\documentclass{ar-1col}
\usepackage[numbers]{natbib}

\jname{Xxxx. Xxx. Xxx. Xxx.}
\jvol{AA}
\jyear{YYYY}
\doi{10.1146/((please add article doi))}

\begin{document}
    \setlength{\pdfpagewidth}{8.5in}%
    \setlength{\pdfpageheight}{11in}%
\markboth{Thomas Vojta}{Disorder in quantum many-body systems}

\title{Disorder in quantum many-body systems}

\author{Thomas Vojta
\affil{Department of Physics, Missouri University of Science and Technology, Rolla, MO 65409, USA; email: vojtat@mst.edu}
}

\begin{abstract}
Impurities, defects, and other types of imperfections are ubiquitous in realistic
quantum many-body systems and essentially unavoidable in solid state materials.
Often, such random disorder is viewed purely negatively as it is believed to prevent
interesting new quantum states of matter from forming and to smear out sharp features
associated with the phase transitions between them. However, disorder is also
responsible for a variety of interesting novel phenomena that do not have clean
counterparts. These include Anderson localization of single particle wave
functions, many-body localization in isolated many-body systems,
exotic quantum critical points, and ``glassy'' ground state phases.
This brief review focuses on two separate but related subtopics in this field.
First, we review under what conditions different types of randomness affect
the stability of symmetry-broken low-temperature phases in quantum many-body systems
and the stability of the corresponding phase transitions.
Second, we discuss the fate of quantum phase transitions that are destabilized by
disorder as well as the unconventional quantum Griffiths phases that emerge in their vicinity.
\end{abstract}

\maketitle

\tableofcontents

\section{INTRODUCTION}

Most real-life quantum many-body systems contain various types of random imperfections
including vacancies, impurity atoms, and extended defects. Such randomness or  disorder
is essentially unavoidable in solid state materials as it arises naturally in the sample preparation process.
Disorder has also been introduced artificially into intrinsically very clean
many-body systems such as ultracold atomic gases in optical lattices.

The effects of disorder on the phases and phase transitions of quantum many-body
systems are often seen in negative terms, a view that Andy Mackenzie
succinctly summarized in the statement: {\it ``For the most part, disorder in condensed matter is a
pain in the neck and a barrier to truth and enlightenment''} \cite{Mackenzie_talk17}.
This perspective stems from the fact that random disorder can suppress
new states of matter,
either by preventing spontaneous symmetry-breaking or by smearing sharp features
in the density of states. Moreover, disorder can round the singularities
associated with phase transitions and critical points.

This review advocates for a more nuanced view: Whereas disorder can indeed do all of these
negative things, it also leads to exciting, qualitatively new phenomena
that do not have clean counterparts. For example, disorder
can induce the spatial localization of the wave function
of a quantum particle, even in the absence of interactions \cite{Anderson58}.
The transition of states at the Fermi energy
from extended to localized behavior is one of the possible mechanisms for metal-insulator
transitions (see, e.g., Refs.\cite{KramerMacKinnon93,EversMirlin08}).
Building on this insight, the combined effects of disorder and interactions on transport
properties have been studied extensively, leading to the identification and analysis of
different universality classes of metal-insulator transitions
\cite{LeeRamakrishnan85,AltshulerAronov85,BelitzKirkpatrick94}.

In recent years, localization in disordered quantum many-body systems has
reattracted enormous attention, albeit in a different context. The field of
\emph{many-body localization} deals with the very foundations of quantum statistical mechanics
by exploring under what conditions an isolated quantum many-body system thermalizes.
Systems that fail to quantum thermalize are many-body localized;  their properties are
not captured by conventional quantum statistical mechanics. Reviews
of this field can be found, e.g., in Refs.\ \cite{NandkishoreHuse15,AltmanVosk15,AbaninPapic17}.

The combination of disorder and interactions can also induce novel low-temperature phases
that are unique to disordered systems. These include, for example, the random-singlet phases in disordered quantum
spin chains \cite{MaDasguptaHu79,DasguptaMa80,Fisher94} as well as various spin glass and electric dipole glass
phases in which the relevant degrees of freedom are frozen in random directions \cite{BinderYoung86,FischerHertz_book91,Mydosh93,VugmeisterGlinchuk90}.

Disorder effects in quantum many-body systems are an enormously broad area that is
impossible to cover in this short review. Instead, we
focus on two separate but related topics, viz., (i) the stability of clean
symmetry-broken low-temperature phases and their quantum phase transitions
against different types of disorder and
(ii) the properties of quantum phase transitions that have been
destabilized by disorder.
We start by reviewing several stability criteria. They were originally derived for
classical systems but have now been established, generalized, and in some cases rigorously proven for
quantum systems at low temperatures.
The corresponding results are scattered throughout the literature; our goal is
to collect them all in one place. In the second part of this article, we review the fate of
quantum phase transitions in disordered systems, and we discuss the exotic quantum
Griffiths phases that emerge in their vicinity.
Parts of the latter material have been reviewed in Refs.\ \cite{Vojta06,Vojta10}.
Here, we therefore emphasize the improved classification of critical points developed in Ref.\
\cite{VojtaHoyos14} that combines and reconciles rare region effects with the Harris criterion.
We also discuss recent experiments.

\section{STABILITY OF PHASES AGAINST DISORDER}
\label{sec:phases}
\subsection{Symmetries and order parameters}
\label{subsec:symmetries}

Landau \cite{Landau37b,Landau37d} developed a general framework for classifying the phases
in macroscopic many-body systems. Different phases can be distinguished according to their
symmetries, and phase transitions generally involve the spontaneous breaking of one or more of
the symmetries of the underlying Hamiltonian.\footnote{Currently, great
research efforts are directed at phases that do not follow Landau's classification but
feature unconventional topological order due to the long-range entanglement of their
quantum wave functions \cite{AletWalczakFisher06,CastelnovoTrebstTroyer10,CastelnovoMoessnerSondhi12,Wen13,Senthil15,SavaryBalents17}.
The study of disorder effects on these phases and their transitions is still in its infancy
and therefore not considered in this article.}
For example, a ferromagnetic phase breaks the global spin rotation symmetry, while the $U(1)$ symmetry
associated with the phase of the macroscopic wave function is broken in a superfluid phase.

In some ordered
 phases, the broken symmetries include real-space symmetries. This is the case,
for instance, in a charge density wave phase that spontaneously breaks the translation and rotation
symmetries of the underlying solid. Other ordered phases, such as the ferromagnetic and superfluid
phases mentioned above, do not break real-space symmetries but only symmetries associated
with spin, phase, or other degrees of freedom. This distinction will become crucial when we
introduce disorder into our system.

To quantify the degree of symmetry breaking, Landau also introduced the concept of order parameters.
An order parameter is a thermodynamic quantity that is zero if the corresponding symmetry is not
broken (i.e., in the disordered\footnote{Unfortunately,
the term ``disorder'' has two different meanings in the field.
On the one hand, ``disordered'' refers to a \emph{state} or phase without
long-range (broken-symmetry or topological) order. The paramagnetic phase of
a magnetic material is called a disordered phase, for example. On the other hand, ``disorder'' denotes
randomness in the underlying \emph{system}, i.e., the Hamiltonian. }
 phase), whereas it is nonzero and usually nonunique in the phase
that breaks the symmetry (the ordered phase). In our example of a ferromagnetic phase,
the total magnetization $\mathbf{m}$ (which is an $O(3)$ vector) is an order parameter.
The order parameter for the superfluid phase is the ``condensate wave function'' $\Psi$, a
complex variable. For charge density wave order with a single allowed wave vector $\mathbf{Q}$,
a complex order parameter $\phi$ can be defined from a Fourier expansion of the
charge density $\rho$ via $\rho(\mathbf{x}) = \rho_0 + \mathbf{Re}(\phi e^{i \mathbf{Q} \cdot \mathbf{x}})$.
If more than one wave vector is allowed, the order parameter becomes a complex vector.

In addition to the general framework for classifying phases, Landau put forward an approximate
quantitative description, the Landau theory of phase transitions. It is based on
an expansion of the free energy density $f$ in powers of all the order parameters in the problem.
In the simplest case of a single scalar order parameter $m$, the Landau expansion reads
$f= -h m + r m^2 +v m^3 + u m^4 + \ldots$ where $h$ is the field conjugate to the order parameter.
The coefficients $r$, $v$, and $u$ can either be treated as phenomenological constants or determined
from a more microscopic calculation. In general, a Landau expansion will contain all terms
that are compatible with the symmetries of the system.

Within Landau theory, the order parameter is a space and time-independent constant. The theory thus
contains neither the spatial inhomogeneities required for describing disorder nor the
order parameter fluctuations necessary to capture the critical behavior near continuous
phase transitions. This can be overcome by considering an order parameter field $m({\mathbf x,\tau})$ that depends on
real space position $\mathbf{x}$ and imaginary time $\tau$.
The Landau free energy gets replaced by the Landau-Ginzburg-Wilson (LGW) free energy functional
\begin{equation}
F = \int_0^\beta d\tau \int d^d x  \left[-h m({\mathbf x,\tau}) + r m^2({\mathbf x,\tau})
+ (\nabla m({\mathbf x,\tau}))^2 + (\partial_\tau m({\mathbf x,\tau}))^2 + \ldots \right]~.
\label{eq:LGW}
\end{equation}
The gradient term punishes rapid changes of the order parameter; it encodes the interactions between
neighboring degrees of freedom. The time derivative term controls the strengths of the quantum fluctuations.
The partition function is now given by a path integral
\begin{equation}
Z = \int D[m({\mathbf x,\tau})]\, \exp\left(- F[m({\mathbf x,\tau})] \right)~.
\label{eq:Z_LGW}
\end{equation}
Equations (\ref{eq:LGW}) and (\ref{eq:Z_LGW}) hold in the quantum case. For classical systems, it
is often sufficient to consider order parameter fields $m(\mathbf{x})$ that depend on space only.
Note that the leading dynamic term in the quantum LGW functional (\ref{eq:LGW}) can take
other forms than $(\partial_\tau m)^2$. Berry phases can produce imaginary terms \cite{Sachdev_book99}.
Moreover, if the system contains soft (gapless) excitations other than
the order parameter fluctuations, the LGW functional generically features nonanalyticities
that stem from integrating out these soft modes
\cite{VBNK96,BelitzKirkpatrickVojta02,BelitzKirkpatrickVojta05}.

\subsection{Types of disorder}

Microscopically, disorder or randomness can have many different origins ranging from impurity atoms and
vacancies to extended defects such as dislocations or grain boundaries in a crystalline solid.
Thin films may experience random strains stemming from a mismatch with the substrate. Almost all
disorder in condensed matter systems is time-independent over the relevant experimental time scales;
this kind of disorder is called \emph{quenched}. In contrast, \emph{annealed} disorder changes over the
time span of a typical experiment. In the present article, we almost exclusively consider quenched disorder.

In Sec.\ \ref{subsec:symmetries}, we have seen that ordered phases can be classified according to which
symmetries they break. This suggests that one should also classify the various types of disorder according
to their symmetries. Consider, for example, an Ising (easy-axis) ferromagnet in an external magnetic field $h(\mathbf{x})$
that varies randomly in space. This type of disorder is called random-field disorder; within a LGW description,
it couples \emph{linearly} to the order parameter. The corresponding term in (\ref{eq:LGW}) reads
\begin{equation}
 -h(\mathbf{x})\, m({\mathbf x,\tau})~.
\label{eq:LGW-RF}
\end{equation}
Random fields locally prefer a particular direction of $m$  and therefore
locally break the spin rotation symmetry. Whether or not it is broken globally
depends on the distribution of $h(\mathbf{x})$. If this distribution is even, the global symmetry
is preserved in a statistical sense because no direction is preferred globally.

Now consider an Ising ferromagnet containing a number of randomly
distributed vacancies. Since the vacancies do not prefer a particular magnetization direction, they do not
break the up-down spin symmetry of the Hamiltonian (neither locally nor globally).
They cause local variations in the tendency towards ferromagnetism, i.e., they change the local critical
temperature. This type of disorder is therefore called random-$T_c$ disorder. Within a LGW
theory, it couples to the square of the order parameter,\footnote{In quantum field theory,
the quadratic term contains the mass of the particle. Random-$T_c$ disorder is thus also called
random mass disorder.}
leading to
a random variation $\delta r (\mathbf{x})$ in space of the quadratic coefficient. The quadratic term now reads
\begin{equation}
 [r+\delta r (\mathbf{x})]\, m^2({\mathbf x,\tau})~.
\label{eq:LGW-RM}
\end{equation}

Many additional kinds of disorder can appear in quantum many-body systems. For example, the disorder can consist
of random phase shifts for a complex order parameter, or it can introduce easy axes in random directions in an
XY or Heisenberg magnet. Moreover, strong disorder can lead to frustrated interactions that can
change the thermodynamic phases qualitatively.

\subsection{Imry-Ma criterion: symmetry-breaking and random-field disorder}
\label{subsec:Imry-Ma}

In this section, we sketch the derivation of a criterion for the stability
of a spontaneously symmetry-broken phase against random-field disorder.
To be specific, consider an Ising ferromagnet subject to uncorrelated random fields
that have a symmetric distribution of zero mean $[h(\mathbf{x})]_{dis}=0$ and variance
$[h(\mathbf{x})h(\mathbf{x}')]_{dis}=W\delta(\mathbf{x}-\mathbf{x}')$.
In this system, the spin ``up-down'' symmetry is locally broken because spatial regions with
positive local field $h$ prefer a positive magnetization
$m$ while regions with negative $h$ prefer a negative $m$.
However, the random fields preserve the global symmetry in a statistical sense.
The central question of this section is:
Is global spontaneous symmetry breaking into a long-range ordered ferromagnetic state (in which
the magnetization is either positive everywhere or negative everywhere)
still possible?

\begin{figure}
\includegraphics[width=3.8in]{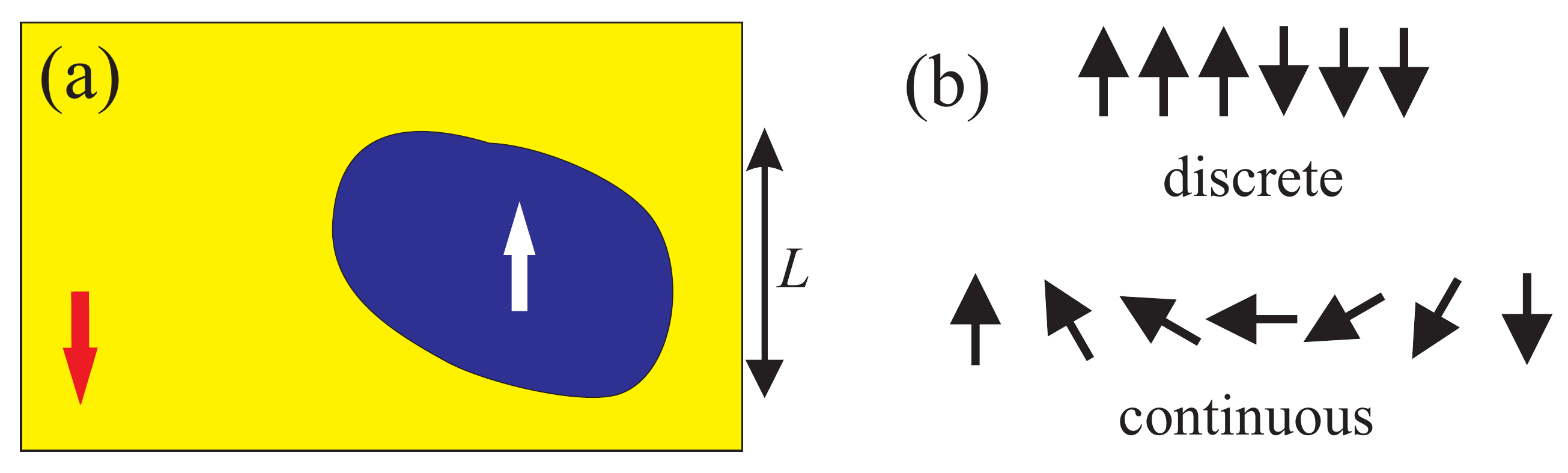}
\caption{(a) The Imry-Ma criterion compares the free energy gain from
aligning a domain of linear size $L$ with the average local random field
to the free energy cost for creating the domain wall.
(b) For discrete order parameter symmetry, the domain wall is sharp, i.e., it has a fixed
width independent of the domain size. For continuous symmetry, the change in
order parameter orientation can be spread out over a length of order $L$.}
\label{fig:Imry-Ma}
\end{figure}

To answer this question, Imry and Ma \cite{ImryMa75} derived a criterion for the stability of
the ferromagnetic state against domain formation. Consider a system in a putative
``spin-down'' ferromagnetic state containing a spatial region of linear size $L$ in
which the average random field is positive and thus prefers a ``spin-up'' order parameter,
as shown in Fig.\ \ref{fig:Imry-Ma}a.
To decide whether a ``spin-up'' domain forms,
one needs to weigh the free energy gain due to aligning the domain with the average local
random field against the free energy cost for the domain wall. In $d$ space dimensions,
the domain wall is a $(d-1)$-dimensional hyper surface; its energy cost can therefore
be estimated as $\Delta F_{DW} \sim \sigma L^{d-1}$ where the constant $\sigma$ is the
surface energy density. The energy gain from aligning the domain with the local
random field is proportional to the integral of $h(\mathbf{x})$ over the domain.
Estimating the typical value of this integral via the central limit theorem leads to
$|\Delta F_{RF}| \sim W^{1/2}L^{d/2}$. The uniform ferromagnetic state is stable if
$|\Delta F_{RF}| <  \Delta F_{DW}$ for all potential domain sizes $L$.

For $d>2$, $\Delta F_{DW}$ grows faster with $L$ than $|\Delta F_{RF}|$. Thus,
domains will not form if the random fields are weak, implying that the ferromagnetic
state is stable. In contrast, for $d<2$, the random field term $|\Delta F_{RF}|$
will overcome the domain wall energy $\Delta F_{DW}$ for sufficiently large $L$
even if the random fields are weak. This means that the uniform ferromagnetic state is
destroyed by domain formation.

Aizenman and Wehr \cite{AizenmanWehr89} later proved
rigorously that random field disorder prevents spontaneous symmetry breaking in
dimensions $d\le 2$ for discrete order parameter symmetry and for $d\le 4$ in the case
of continuous symmetry. The continuous symmetry case is different
because the domain wall can be spread out over the entire domain (see Fig.\ \ref{fig:Imry-Ma}b).
A simple estimate of the gradient term in the LGW functional (\ref{eq:LGW})
yields $\Delta F_{DW} \sim L^d (\nabla m)^2 \sim L^{d-2}$ which results in a critical dimension
of 4. So far, we have considered uncorrelated random fields.
Long-range correlated random fields with correlations that decay as $|\mathbf{x}-\mathbf{x}'|^{-a}$
have stronger effects if $a<d$. In this case, domain formation is favored for
$a<2$ whereas the uniform ferromagnetic state is stable for $a>2$ \cite{Nattermann83}.

The Imry-Ma criterion shows that arbitrarily weak random fields prevent spontaneous symmetry
breaking in $d\le 2$. However, the length scale beyond which domains destroy the uniform state,
the so-called breakup length $L_B$, depends sensitively on the random field strength.
Comparing $|\Delta F_{RF}|$ and $\Delta F_{DW}$ yields $L_B \sim (W/\sigma^2)^{1/(d-2)}$.
For the marginal dimension $d=2$, the dependence becomes exponential,
$L_B \sim \exp(\textrm{const}/W)$, implying that domains become important only at very large scales
for weak random fields.

Although the Imry-Ma criterion was originally derived for classical systems,
it also applies to quantum systems at low or zero temperature. This stems from the fact
that the disorder varies only in space but not in (imaginary) time. A quantum version
of the rigorous Aizenman-Wehr theorem was recently proven by Greenblatt et al.
\cite{GreenblattAizenmanLebowitz09,AizenmanGreenblattLebowitz12}.

\subsection{When do random fields emerge?}

How does random-field disorder arise in realistic quantum many-body systems?
To answer this question, it is crucial to distinguish order parameters that break
real-space symmetries from order parameters that only break symmetries that do
not involve real space.

If an order parameter does not break real-space symmetries,  generic disorder
does not produce random fields because it does not locally break the order parameter
symmetry.
For example, vacancies in a ferromagnet do not break the
spin rotation symmetry. Analogously, disorder in the Josephson couplings
in a Josephson junction array does not break the $U(1)$ symmetry of the
superfluid order parameter. This means that the disorder
does not couple to the order parameter $m$ linearly in an LGW theory,  Instead, it generically couples
to $m^2$, i.e., it acts as random-$T_c$ disorder.

In contrast, for order parameters that break real-space symmetries, vacancies,
impurities and other defects generically generate random fields because
they locally break the corresponding symmetries. For example, an electronic nematic
phase spontaneously breaks the rotation symmetry of the underlying crystal lattice
\cite{KivelsonFradkinEmery98,FKLEM10,FernandesChubukovSchmalian14}. Local arrangements
of impurities will generally prefer a particular orientation of the nematic
order, breaking its symmetry locally. They thus
act as random fields and couple linearly to the order parameter in a LGW theory
\cite{CDFK06}.
Analogously, a charge density wave spontaneously breaks
the translational symmetry. Impurities generally prefer regions of either
low or high density, i.e., a particular phase of the charge density wave.
Consequently, they act as random field disorder which destroys the
charge density wave phase for $d\le 4$.

Instead, the disorder induces an exotic ``Bragg glass'' with power-law correlations
(in $d=3$ and for weak disorder) \cite{GiamarchiLeDoussal94,GingrasHuse96,Fisher97}.
It has been observed, for example, in the vortex lattice of a type II superconductor
\cite{Kleinetal01}. Recently, similar
spin-density-wave and pair-density-wave glass phases have been discovered
in situations where long-range spin-density-wave or pair-density-wave order
is destroyed by impurities \cite{MrossSenthil15}.

Random fields can also arise via more subtle mechanisms. LiHoF$_4$ is a dipolar
Ising magnet. A magnetic field applied perpendicular to the Ising axis suppresses
$T_c$ and induces a quantum phase transition
to a paramagnetic state \cite{BitkoRosenbaumAeppli96}. If the magnetic Ho ions
are replaced by nonmagnetic Y ions in LiHo$_{1-x}$Y$_x$F$_4$, the interplay between
the dilution, the off-diagonal terms of the dipolar interaction, and the applied transverse
field (which breaks time-reversal symmetry) generates longitudinal random fields
that qualitatively change the low-temperature behavior
\cite{TGKSF06,SBBGAR07,Schechter08,GingrasHenelius11}.

\subsection{Example: random-field disorder from vacancies}
\begin{figure}
\includegraphics[width=5in]{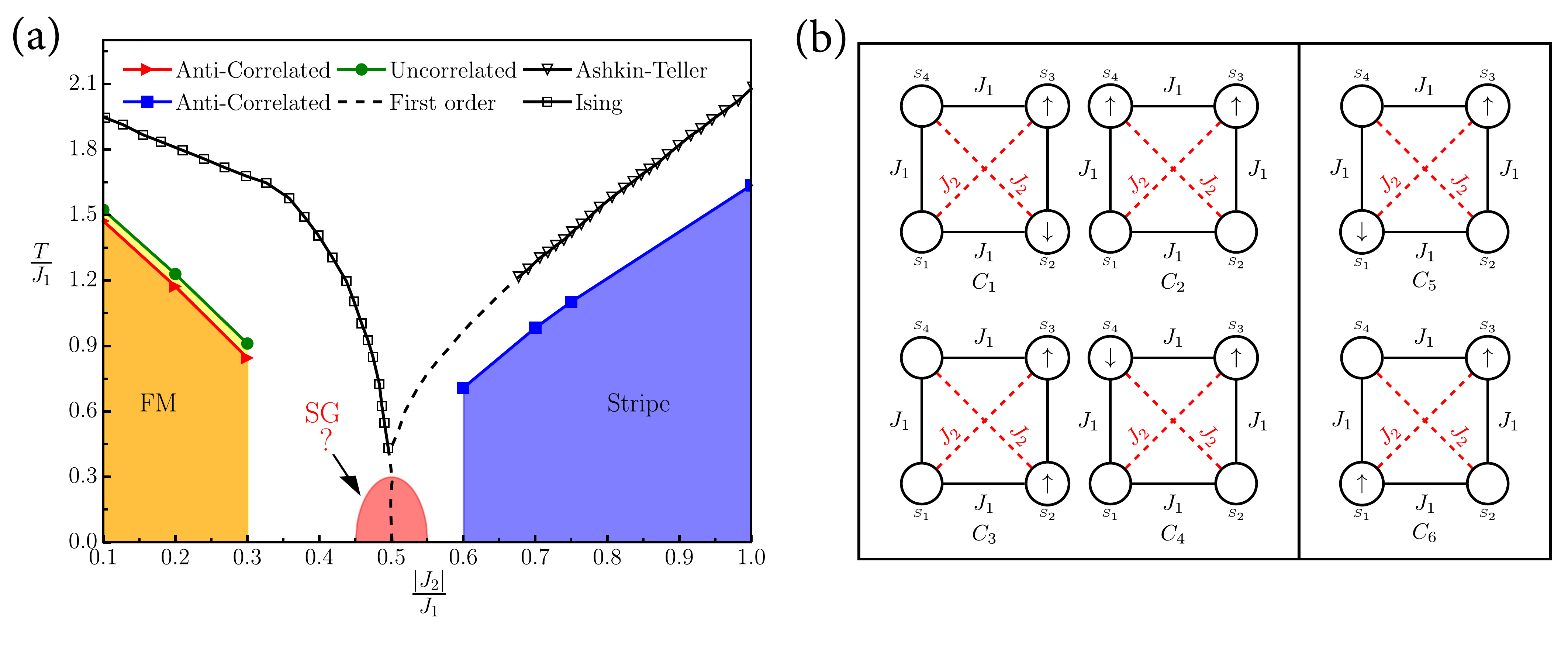}
\caption{(a) Phase diagram of the $J_1$-$J_2$ model (\ref{eq:H_J1J2})
for uncorrelated and anticorrelated vacancies of concentrations $p=1/8$
as well as the undiluted system (open symbols). (b) Impurity configurations
on $2\times 2$ plaquettes illustrating the random-field mechanism
(from Ref.\ \cite{KunwarSenVojtaNarayanan18}).
}
\label{fig:J1J2}
\end{figure}

The diluted frustrated square-lattice Ising model with ferromagnetic nearest-neighbor
interactions $J_1>0$ and antiferromagnetic next-nearest-neighbor interactions $J_2<0$
is given by
\begin{equation}
H = -J_{1}\sum_{\langle ij \rangle }\rho_{i}\rho_{j} S_{i}S_{j} - J_{2} \sum_{\langle \langle ij \rangle \rangle}\rho_{i}\rho_{j}S_{i}S_{j}~.
\label{eq:H_J1J2}
\end{equation}
$S_{i} =\pm 1$ is an Ising spin, and the random variable $\rho_i$ takes values
0 (vacancy) or  1 (occupied site) with probabilities $p$ and $1-p$, respectively.
The undiluted system features two distinct symmetry-broken phases
(see Fig.\ \ref{fig:J1J2}a).
For $|J_2|/J_1 < 1/2$, the low-temperature phase is ferromagnetic,
but for $|J_2|/J_1 > 1/2$, the system displays stripe order
characterized by a two-component order parameter
$\psi_x = (1/L^2)\sum_{i}\rho_{i}S_i(-1)^{x_i}$, $\psi_y = (1/L^2)\sum_{i=1}\rho_{i}S_i(-1)^{y_i}$
where $x_i$, $y_i$ are the coordinates of site $i$.
The ferromagnetic phase breaks just the $Z_2$ Ising symmetry, but the stripe phase also breaks the $Z_4$
lattice rotation symmetry.

Since spinless impurities do not break the Ising symmetry, they
do not create random fields for the ferromagnetic order parameter, the magnetization $m$. Instead
they act as random-$T_c$ disorder and couple to $m^2$. Consequently, the ferromagnetic phase is expected
to survive in the presence of impurities.

Even though a single impurity does not break the $Z_4$ lattice rotation symmetry, spatial arrangements
of more than one impurity do. If two vertical nearest neighbors are both occupied by impurities,
vertical stripes have a lower energy (by $-2J_1$) than horizontal stripes (see Fig.\ \ref{fig:J1J2}b).
Similarly, if impurities occupy two horizontal nearest neighbors, horizontal stripes are favored.
Impurities on nearest neighbor sites thus create random fields
for the nematic order parameter $\eta=\psi_x^2-\psi_y^2$ that are expected to destroy
the stripe phase.

Monte Carlo simulations with uncorrelated impurities
\cite{KunwarSenVojtaNarayanan18}
have confirmed that the stripe phase is destroyed while the ferromagnetic phase survives
(see Fig.\ \ref{fig:J1J2}a). A similar mechanism was identified in
an XY antiferromagnet on a pyrochlore lattice \cite{AndradeHoyosRachelVojta18}.

Because random fields only appear if pairs of impurities occupy nearest neighbor sites,
they will be absent for perfectly anticorrelated impurities
where such pairs are forbidden.\footnote{At zero temperature, this is an exact result.
Entropic effects may generate random fields at nonzero temperatures \cite{SVSN93},
but they are expected to be extremely weak at low temperatures.}
Monte Carlo simulations
\cite{KunwarSenVojtaNarayanan18} indeed show that the stripe phase survives the introduction of
perfectly anticorrelated disorder, see Fig.\ \ref{fig:J1J2}a.
The preservation of the stripe phase by anticorrelations between impurities
is analogous to the protection of clean quantum critical points by local disorder
correlations in a random quantum Ising chain \cite{HLVV11}.

\section{STABILITY OF PHASE TRANSITIONS AGAINST DISORDER}

We now turn to the stability of phase transitions against disorder.
The focus will be on random-$T_c$ disorder
because random-field disorder completely prevents symmetry-breaking in $d\le 2$.
If the ordered phase
survives in the presence of random fields in $d>2$, its phase transition is usually
controlled by a classical zero-temperature renormalization group fixed point
\cite{BrayMoore85}. This means that the static random-field fluctuations
dominate over both the thermal fluctuations and the quantum fluctuations.
This has been explicitly demonstrated, for example, for the quantum spherical
model \cite{Vojta96} in a random field \cite{VojtaSchreiber96}.

In contrast, weak random-$T_c$ (random-mass) disorder does not affect the stability
of the bulk phases, but it can destabilize the phase transitions between them. In this
section, we review the corresponding stability criteria.

\subsection{Imry-Ma criterion again: stability of first-order transitions against random-$T_c$ disorder}

First order phase transitions are characterized by the macroscopic coexistence
of two distinct phases at the transition point. Random-$T_c$ disorder
locally favors one phase over the other. We therefore arrive at the same question as
in Sec.\ \ref{subsec:Imry-Ma}: Will uniform macroscopic phases survive at the
transition point or will the system form finite-size domains of the locally favored
phase?

To answer this question, one can adapt the Imry-Ma criterion
\cite{ImryWortis79,HuiBerker89}. Consider a single domain of the first
phase located in a favorable region of the random-$T_c$ disorder and
embedded in the second phase. The free energy cost of the surface
increases as $\Delta F_{surf} \sim \sigma L^{d-1}$ with domain size
$L$ where $\sigma$ is the surface energy density.\footnote{This terms scales as $L^{d-1}$ independent of the symmetry
of the order parameter within each phase because the two distinct phases are
generally \emph{not} connected via a continuous transformation. }
The free energy gain of the domain from being in the ``right'' phase is obtained
from central limit theorem as $|\Delta F_{dis}| \sim W^{1/2} L^{d/2}$
where $W$ is the variance of the random $T_c$ disorder.
Phase coexistence is therefore impossible in $d\le 2$ for arbitrarily weak
random $T_c$ disorder. This means the first-order phase transition is destroyed.
For $d>2$, phase coexistence is possible, and the first-order transition
survives for disorder strengths below a certain threshold.

Since all of these results had originally been derived for classical phase transitions,
there was some uncertainty initially about their applicability to quantum
phase transitions \cite{GoswamiSchwabChakravarty08}. However, a quantum version
of the Aizenman-Wehr theorem has now been proven
\cite{GreenblattAizenmanLebowitz09,AizenmanGreenblattLebowitz12}. Moreover,
explicit results for first-order quantum phase transitions
in various types of quantum spin chains confirm the criterion
\cite{GoswamiSchwabChakravarty08,SenthilMajumdar96,HrahshehHoyosVojta12,Barghathietal14}.

The question of what happens to a first-order transition that is
destabilized by random-$T_c$ disorder is beyond the reach of the Imry-Ma criterion.
Transitions between an ordered and a disordered phase are often rounded into
continuous ones. The fate of transitions between two different
ordered phases is more complex because Landau's classification does not
allow such transitions to be continuous.\footnote{Continuous phase transitions between different
ordered phases can occur within exotic scenarios such as
deconfined quantum criticality  \cite{SVBSF04,SBSVF04}.}
Therefore, an intermediate phase often appears.

\subsection{Harris criterion: stability of critical points}

To derive a criterion for the stability of a clean
critical point against weak random-$T_c$ disorder, we divide the
system into blocks whose size is the correlation length $\xi$.
Because of the disorder, each block $i$ has its own critical temperature $T_c(i)$.
We now compare the variations $\Delta T_c$
of these block critical temperatures with the distance $T-T_c$ from the global
critical point. As long as $\Delta T_c < |T-T_c|$, all blocks are in the same phase,
and the system is approximately uniform.  For $\Delta T_c > |T-T_c|$, however,
different blocks are on different sides of $T_c$, making a uniform transition impossible.

Consequently, the clean critical behavior is stable if $\Delta T_c < |T-T_c|$
remains valid as the transition is approached, i.e., for $\xi \to \infty$.
Because the $T_c(i)$ of a block is determined by the average over a large number of
random variables, central limit theorem predicts that
$\Delta T_c \sim \xi^{-d/2}$. The global distance from criticality
is related to the correlation length via $\xi \sim |T-T_c|^{-\nu}$
where $\nu$ is the clean correlation length exponent. The condition $\Delta T_c < |T-T_c|$
in the limit $\xi \to \infty$ implies Harris' exponent inequality \cite{Harris74}
\begin{equation}
d \nu  > 2~.
\label{eq:Harris}
\end{equation}
If Harris' inequality is fulfilled, the ratio $\Delta T_c / |T-T_c|$ approaches
zero for $\xi\to \infty$. The system thus becomes asymptotically clean at large length scales.
In contrast, if Harris' inequality is violated, $\Delta T_c / |T-T_c|$ increases
as the transition is approached, destabilizing the uniform clean
transition. We emphasize
that the Harris criterion is a necessary condition for the stability of the clean
critical point, not a sufficient one because it only tests the
self-consistency of the clean behavior in the large length-scale limit.
New physics that the disorder may induce at finite scales is invisible to the Harris
criterion.

Just as the Imry-Ma criterion, the Harris criterion
(\ref{eq:Harris}) was originally derived for classical phase transitions.
It takes the same form for zero-temperature quantum phase transitions because
quenched disorder varies only in space but not in (imaginary) time. (The
dimensionality $d$ in Harris' inequality (\ref{eq:Harris}) is \emph{not} replaced
by $d+1$  or $d+z$ in the quantum case.)

Harris' original criterion (\ref{eq:Harris}) which applies to
uncorrelated spatial disorder has been generalized in several directions.
For extended defects, i.e., disorder perfectly correlated
in at least one space dimension, the inequality reads $d_\perp \nu >2$ where $d_\perp$
is the number of dimensions in which there is randomness ($d_\perp = d-1$ for line defects
and $d_\perp=d-2$ for plane defects). If the disorder features isotropic long-range correlations in space
that decay as  $|\mathbf{x}-\mathbf{x}'|^{-a}$,
the Harris criterion is modified to $\min(d,a) \nu >2$, making
long-range correlated disorder with $a<d$ more relevant
than uncorrelated disorder \cite{WeinribHalperin83}.
Harris-like criteria can also be derived for disorder that varies in time or in space and time.
For purely time-dependent disorder with short-range correlations, the resulting
inequality reads $z\nu > 2$ where $z$ is the dynamical critical exponent
\cite{Kinzel85,AlonsoMunoz01}. Recently, Vojta and Dickman derived a
criterion for arbitrary spatio-temporal disorder in terms of its space-time
correlation function \cite{VojtaDickman16}. It contains the older results as special
cases but also works for more complicated situations such as
diffusive disorder degrees of freedom.

Another generalization of the Harris criterion is due to Luck \cite{Luck93a}
who considered the stability of critical points not just against random disorder but against
a broader class of inhomogeneities whose fluctuations can be characterized by a wandering
exponent $\omega$. In terms of this exponent, the stability criterion reads
$\omega < 1 - 1/(d\nu)$. The Harris-Luck criterion has been used, for example, for systems
with quasiperiodic inhomogeneities.

Violations of the Harris criterion are sometimes reported in the literature,
for example, for phase transitions on random Voronoi lattices (see Ref.\
\cite{BarghathiVojta14} and references therein) or in certain dimerized spin
models \cite{YaoGustafssonCarlsonSandvik10,MaSandvikYao14}. In the former
case, they stem from hidden anticorrelations of the disorder variables
caused by a topological constraint \cite{BarghathiVojta14}. The violations
in the latter systems have been attributed to the fact that the disorder
causes no (or extremely small) shifts of the local transition point.

Finally, we emphasize that the Harris criterion (\ref{eq:Harris})
tests the stability of the \emph{clean}
critical point and contains the correlation length exponent $\nu$
of the \emph{clean} system.
The separate question which value $\nu$ takes at the disordered critical point was addressed
by Chayes et al. \cite{CCFS86} who showed that the finite-size correlation
length exponent in a disordered system must fulfill the same inequality
$d\nu \ge 2$. However, there are unresolved questions about the relation
between the finite-size correlation length exponent and the intrinsic one
\cite{PazmandiScalettarZimanyi97}.

\section{DISORDERED PHASE TRANSITIONS}
\label{sec:QPT}

So far, we have discussed the stability of clean phases and phase transitions against
(weak) random-field and random-$T_c$ disorder. We now turn to the ultimate fate of
a transition in the presence of disorder.
The focus will be on critical points because first-order
phase transitions cannot exist in disordered system for $d\le 2$, and comparatively
little is known about disordered first-order (quantum) phase transitions in $d>2$.
Parts of this topic have been reviewed recently in Refs.\
\cite{Vojta06,Vojta10,Vojta13,Vojta14}. We therefore only summarize the key concepts
and emphasize the refined classification developed in Ref.\ \cite{VojtaHoyos14}.

\subsection{Clean vs.\ finite-disorder vs.\ infinite-disorder critical points}
\label{subsec:clean_vs_disordered}

Critical points in disordered systems can be categorized according to the behavior
of the disorder strength under coarse graining \cite{MMHF00}. Three cases can be
distinguished:

(i) If the Harris criterion is fulfilled, the disorder strength goes to
zero under coarse graining, i.e., disorder is irrelevant in the renormalization group
sense. The resulting critical behavior equals that of the clean transition,
and macroscopic observables are self-averaging.

(ii) The second case comprises critical points at which
the system remains inhomogeneous, and the (relative) strength
of the disorder approaches a nonzero constant in the large length scale limit.
These ``finite-disorder'' critical points generally show conventional power-law
critical behavior, but the critical exponents differ from the corresponding clean ones.
Macroscopic observables are not
self-averaging at criticality; their distribution retains a finite
width in the thermodynamic limit \cite{WisemanDomany98}.

\begin{textbox}
\section{Marginal case: the example of the 2d Ising universality class}
The correlation length exponent $\nu=1$ of the 2d Ising universality class is
exactly marginal w.r.t.\ the Harris criterion, $d\nu=2$. Is random-$T_c$ disorder
relevant or irrelevant? The critical behavior of a 2d disordered Ising magnet has been
controversially discussed for a long time, but recent high-accuracy Monte
Carlo simulations \cite{ZWNHV15} provide strong evidence in favor of the
strong-universality scenario  \cite{DotsenkoDotsenko83,Shalaev84,Shankar87}
according to which the critical behavior is controlled by the clean
Ising fixed point. Disorder is  marginally \emph{irrelevant} and gives rise to
universal logarithmic corrections to scaling (see Ref.\ \cite{ZWNHV15} and
references therein). Interestingly, the same clean Ising behavior
with logarithmic corrections also governs the critical point of the disordered
$N$-color Ashkin-Teller model that emerges when the clean first-order transition
is destroyed by disorder \cite{ZWNHV15,Cardy96,Cardy99}.
\end{textbox}

(iii) In the third case, the disorder strength (the relative magnitude of the
inhomogeneities) goes to infinity in the limit of large length scales.
The resulting infinite-disorder (or infinite-randomness) critical points
usually show unconventional activated scaling behavior \cite{Fisher92,Fisher95}
featuring an exponential relation between correlation length and time rather than
the usual power-law relation.

\subsection{Rare regions and Griffiths singularities}
\label{subsec:RR}

Recent research has shown that many phase transitions in disordered systems
are dominated by rare strong disorder fluctuations and the rare spatial
regions on which they reside. Rare regions cause off-critical singularities
in the free energy called the Griffiths singularities \cite{Griffiths69}.

\begin{figure}
\includegraphics[width=3.8in]{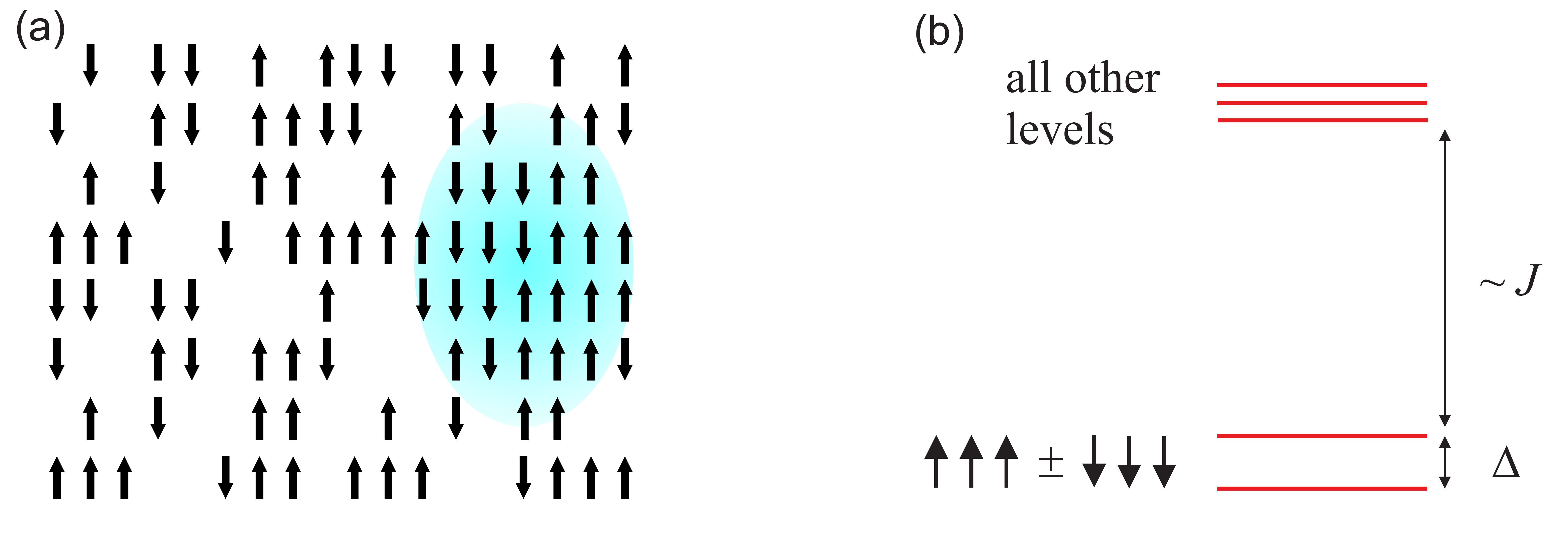}
\caption{(a) Rare region in a diluted ferromagnet. The shaded region is impurity-free
and thus behaves as a finite-size piece of the undiluted system.
(b) Energy spectrum of a single rare region in a quantum Ising magnet.
In the two low-energy states, all spins on the rare region are aligned.
They are separated from all other states by a large gap of the order of
the interaction energy $J$. }
\label{fig:RR}
\end{figure}

The importance of rare regions can be discussed using the example of a diluted ferromagnet
shown in Fig.\ \ref{fig:RR}a.
Due to statistical fluctuations of the vacancy positions,
a macroscopic sample contains a small but nonzero concentration of large
vacancy-free regions. If the system as a whole is close to
the transition but still on the paramagnetic side, such regions can be locally
ferromagnetic, i.e., their spins lock together and align parallel.

To decide whether or not rare regions play a significant role,
one must estimate their total contribution to thermodynamic quantities.
The probability for finding a large vacancy-free region of size $L_{RR}$
is exponentially small
in its volume $V_{RR}\sim L_{RR}^d$ and in the vacancy concentration $c$. Up to pre-exponential
factors it reads $w(V_{RR}) \sim \exp(-cV_{RR})$.\footnote{This estimate
holds for uncorrelated disorder. If the disorder features
long-range correlations that decay as $|\mathbf{x}-\mathbf{x}'|^{-a}$ with $a<d$,
rare regions are much more likely to occur. Their probability is enhanced and reads
$w(V_{RR}) \sim \exp(-c V_{RR}^{a/d})$ \cite{IbrahimBarghathiVojta14}.}
Consequently, rare regions are important only if the contribution each
one makes increases exponentially
with its volume. At generic classical transitions, this is not the case.
Each locally ordered region in a diluted ferromagnet, for example,
acts as a superspin whose moment is proportional to the volume $V_{RR}$.
The susceptibility of the rare region thus behaves as
$\chi(V_{RR}) \sim V_{RR}^2/T$. As this power-law increase cannot overcome the
exponential decrease of the rare region probability with $V_{RR}$,
large rare regions do not make significant contributions.
Thermodynamic Griffiths singularities in generic classical systems are
thus weak essential singularities
that are likely unobservable in experiments
\cite{Wortis74,Harris75,Imry77}.

Quantum systems at zero temperature can have stronger Griffiths singularities.
Consider, for instance, the energy spectrum of a rare region in a diluted Ising
magnet in a transverse magnetic field, as sketched in Fig. \ref{fig:RR}b
\cite{SenthilSachdev96}. The two low-lying states are the symmetric and antisymmetric
combinations of the perfectly aligned ``superspin'' states. They
 are separated by an  exponentially small gap
 $\Delta \sim \exp(-a V_{RR})$, leading to to an exponential
increase of the rare region magnetic susceptibility with $V_{RR}$.
The Griffiths singularities
are therefore much stronger and of power-law from \cite{ThillHuse95,YoungRieger96}.
Power-law Griffiths singularities can also appear at classical transitions
in systems with extended defects, i.e., if the disorder is perfectly correlated in
at least one dimension \cite{McCoyWu68}.

Even stronger rare region effects occur if the dynamics of an individual
rare region can freeze independently of the bulk system. At $T\ne 0$, this can
happen if the disorder is correlated in at least two dimensions
\cite{Vojta03b,SknepnekVojta04}. At quantum phase transitions it can also be caused by
the coupling of the order parameter
to a dissipative bath \cite{MillisMorrSchmalian01,Vojta03a,HoyosVojta08}.

\subsection{Classification of disordered critical points}
\label{subsec:classification}

Vojta and Hoyos \cite{VojtaHoyos14} recently showed that there is a deep connection
between the Harris criterion and rare region physics. This allowed them to
combine the two ways of categorizing critical points introduced in
Secs.\ \ref{subsec:clean_vs_disordered} and \ref{subsec:RR}, leading to an
improved classification
scheme for classical, quantum, and nonequilibrium critical points under the influence of random-$T_c$ disorder
that extends earlier work \cite{Vojta06,VojtaSchmalian05}.
The three main classes are determined by the relation of the effective rare region
dimensionality $d_{RR}$ with the lower critical dimension $d_c^-$ of the phase
transition.\footnote{The lower critical dimension $d_c^-$ is the dimension below which the
ordered phase is destroyed by fluctuations.}
For quantum phase transitions, the effective dimensionality includes
the imaginary time direction as one of the dimensions.

{\bf Class A:} If the dimensionality of the rare regions is below the lower critical dimension,
$d_{RR}<d_c^-$, individual rare regions cannot order by themselves. Their contribution to thermodynamic
observables grows at most as a power of their volume which cannot overcome the exponential
decrease of their probability $w(V_{RR})$. Rare regions therefore make a negligible
contribution to the critical thermodynamics. Transitions in this class include generic
thermal (classical) transitions with uncorrelated disorder ($d_{RR}=0$).
Some quantum phase transitions also belong to this class, such as the transition
in the diluted bilayer antiferromagnet \cite{VajkGreven02,SknepnekVojtaVojta04,VojtaSknepnek06}
or the superfluid-Mott glass transition
\cite{ProkofevSvistunov04,IyerPekkerRefael12,Vojtaetal16}. Here, $d_{RR}=1$
because the disorder is perfectly correlated in imaginary time but
$d_c^-=2$ because of the Mermin-Wagner theorem \cite{MerminWagner66}.
Class A contains two subclasses depending on the Harris criterion. In subclass A1, the disorder strength
asymptotically scales to zero, leading to clean critical behavior. Subclass A2 contains finite-disorder
critical points with conventional power-law scaling but exponents that differ from the clean ones.
For quantum phase transitions, this implies that the dynamical critical exponent $z$ remains
finite.

{\bf Class B:} In this class, the rare regions are right at the lower critical
dimension, $d_{RR}=d_c^-$, but still cannot undergo the transition by themselves.
The rare region contribution to thermodynamic quantities now increases exponentially
with their volume, compensating for the exponential decrease of the rare region probability.
This leads to strong power-law Griffiths singularities controlled by a non-universal Griffiths
dynamical exponent $z'$.

Class B is also divided into two subclasses according to the Harris criterion.
In subclass B1, the disorder strength scales to zero for large length scales. Power-law Griffiths
singularities coexist with clean critical behavior, and the Griffiths dynamical exponent
$z'$ does not diverge but approaches the clean $z$.
Such behavior was recently found at a nonequilibrium transition
\cite{VojtaIgoHoyos14}; it may also explain the stability of the Belitz-Kirkpatrick
critical behavior in weakly disordered metallic ferromagnets
(see Refs.\ \cite{BelitzKirkpatrickVojta05,BrandoBelitzGroscheKirkpatrick16}).
In subclass B2, the disorder strength diverges for large length scales, giving
rise to infinite-disorder criticality with activated scaling ($z$ is formally infinite).
Examples include thermal transitions in systems with extended defects
such as the McCoy-Wu model \cite{McCoyWu68} ($d_{RR}=d_c^-=1$) or Heisenberg magnets with plane
defects \cite{MohanNarayananVojta10,HrahshehBarghathiVojta11} ($d_{RR}=d_c^-=2$). Subclass B2
also contains the quantum phase transitions in the random transverse-field Ising model
\cite{Fisher92,Fisher95,YoungRieger96, MMHF00}, metallic Heisenberg magnets
\cite{HoyosKotabageVojta07,VojtaKotabageHoyos09}, and superconducting nanowires
\cite{HoyosKotabageVojta07,DRMS08,DRHV10}. Disordered absorbing-state transitions also belong to this subclass
\cite{HooyberghsIgloiVanderzande03,VojtaDickison05,VojtaLee06,VojtaFarquharMast09,Vojta12}.

{\bf Class C:} In class C with $d_{RR} > d_c^-$, individual rare regions can undergo the phase transition independently
of the bulk system. The global phase transition is smeared because a nonzero global
order parameter arises as superposition of many independent rare regions, each with its
own transition point.  As the spatial correlation length
does not diverge in this scenario, the Harris criterion does not play a qualitative role.
Smeared classical phase transitions have been discovered in randomly layered Ising magnets
\cite{Vojta03b,SknepnekVojta04} ($d_{RR}=2$ and $d_c^-=1$). Smeared quantum phase transition include
those in metallic Ising magnets \cite{Vojta03a} in the dissipative transverse-field Ising model
\cite{HoyosVojta08,SchehrRieger06}. Absorbing state transitions with extended defects also fall into this class
\cite{Vojta04,DickisonVojta05}.

This classification, summarized in Table \ref{tab:classification},
applies to continuous transitions with random-$T_c$ (random mass)
disorder and sufficiently short-ranged interactions.
\begin{table}
\renewcommand{\arraystretch}{1.2}
\tabcolsep7.5pt
\caption{Classification of critical points in the presence of random-$T_c$ disorder according to the
Harris criterion $d\nu > 2$ and the relation between the rare region dimensionality $d_{RR}$ and the lower
critical dimension $d_c^-$ (after Ref.\ \cite{VojtaHoyos14}).}
\label{tab:classification}
\begin{center}
\hspace*{-3cm}
\begin{tabular}{ccccccc}
\hline\hline
 Class &{ RR dimension}  &  Subclass & Harris criterion & {Griffiths Singularities}   & {Critical behavior} \\
\hline
                            &                                                & A1 & $d\nu>2$  & weak exponential     & clean   \\
\raisebox{1.8ex}[-1.8ex]{A} & \raisebox{1.8ex}[-1.8ex]{$d_{\rm RR} < d_c^-$} & A2 & $d\nu<2$  & weak exponential &  convent. finite disorder\\
\hline

                            &                                                & B1 & $d\nu>2$  & power law, $z'$ remains finite & clean\\
\raisebox{1.8ex}[-1.8ex]{B}& \raisebox{1.8ex}[-1.8ex]{$d_{\rm RR} = d_c^-$}  & B2 & $d\nu<2$  & power law, $z'$ diverges    &   infinite disorder \\
\hline
                          C&                           $d_{\rm RR} > d_c^-$  &    &          & rare regions freeze         &   smeared transition\\
\hline\hline
\end{tabular}
\end{center}
\end{table}
It assumes that the coupling between
the rare regions can be neglected. Long-range interactions such as the RKKY interaction in
metals may thus lead to modifications \cite{DobrosavljevicMiranda05}.

\section{QUANTUM GRIFFITHS PHASES}
\label{sec:QGP}

\begin{figure}
\includegraphics[width=3.1in]{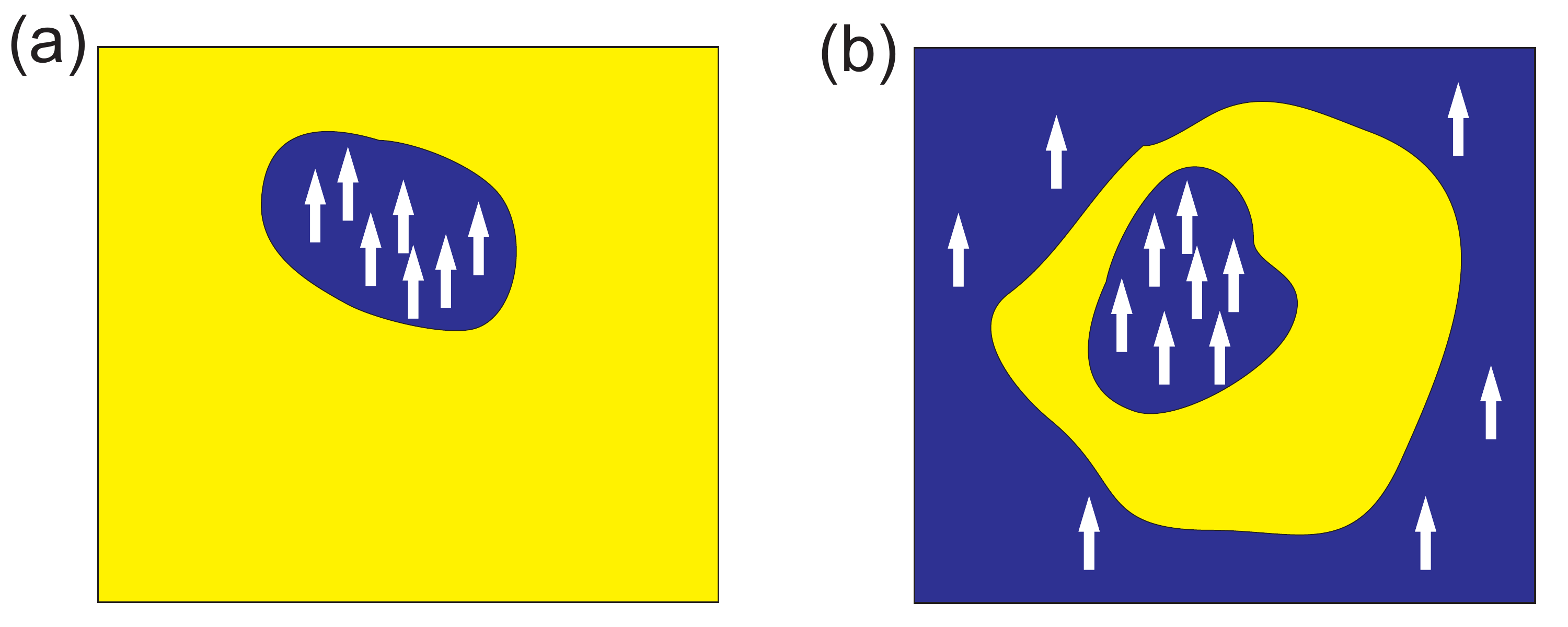}
\caption{Schematic of the rare regions responsible for the Griffiths phases associated with
a ferromagnetic transition. Left: The paramagnetic Griffiths phase is due to rare locally ferromagnetic regions embedded
  in the paramagnetic bulk. Right: The ferromagnetic Griffiths phase is caused by locally ferromagnetic
  regions located inside paramagnetic ``holes'' in the bulk ferromagnet.}
\label{fig:rare_regions}
\end{figure}

In broad terms, a Griffiths phase is a region in the phase diagram of a disordered
system in which the randomness causes finite-size spatial regions to be locally in
the wrong phase. Griffiths phases can appear on both sides of a phase transition;
this is illustrated for a ferromagnet in Fig.\ \ref{fig:rare_regions}.
In the paramagnetic (disordered) Griffiths phase, locally ordered rare regions are embedded
in the paramagnetic bulk system. In the ferromagnetic (ordered) Griffiths phase, in contrast,
the bulk system displays long-range order. The rare regions are not simply holes
in the magnetic order because these holes do not have an associated degree of freedom. Instead, they
are locally ordered clusters inside the holes.\footnote{Are Griffiths phases
distinct phases or just parameter regions within a phase? From a symmetry
perspective, a Griffiths phase is indistinguishable from its parent.
A paramagnetic Griffiths phase, for instance, has the same symmetries as a conventional paramagnet. However,
other qualitative features differ, for example, Griffiths phases are gapless
even if their parent phases are gapped. }

Griffiths phases generically appear close to all phase transitions in disordered
many-body systems, be they thermal, quantum or non-equilibrium transitions.
Here, we focus on quantum Griffiths phases, i.e.,
Griffiths phases that are associated with zero-temperature quantum critical points.
The phenomenology of a quantum Griffiths phase crucially depends on which class of
the classification in Sec.\
\ref{subsec:classification} the critical point belongs to.

For transitions in class A, the rare region density of states decays exponentially
at small energies. Consequently, rare region contributions to thermodynamic quantities
are exponentially suppressed. A prototypical example of a quantum Griffiths phase
in this class is the Mott glass phase emerging in systems of disordered bosons with particle-hole symmetry
\cite{ProkofevSvistunov04,Vojtaetal16,RoscildeHaas07,WeichmanMukhopadhyay08}.
The Mott glass consists of superfluid ``puddles'' embedded in an insulating host;
it is an incompressible insulator just like the conventional Mott insulator.
Whereas the Mott insulator is gapped, the Mott glass is gapless, but
with an exponentially small density of states at low energies.
Rare regions contributions cause
the compressibility to vanish as a stretched exponential with temperature,
$\kappa \sim \exp (-\textrm{const}/T^{1/2})$, i.e., much slower than the
conventional behavior  $\kappa \sim \exp (-\textrm{const}/T)$. This behavior
is an example of an essential Griffiths singularity, as is typical for class A.

Quantum Griffiths phases in class B feature much stronger Griffiths singularities
because the combination of the exponentially decreasing rare region probability and
the exponential dependence of their energy gap (or inverse characteristic time)
on their size leads to a
power-law density of states $g(\epsilon) \sim \epsilon^{d/z'-1}$
that is controlled by the nonuniversal Griffiths dynamic exponent $z'$.
The resulting power-law quantum Griffiths singularities were first found
in random transverse-field Ising models \cite{Fisher92,Fisher95,YoungRieger96, MMHF00}.
Later, they were also predicted to occur in disordered itinerant Heisenberg magnets
\cite{HoyosKotabageVojta07,VojtaKotabageHoyos09} and near the pairbreaking
superconductor-metal quantum phase transition \cite{HoyosKotabageVojta07,DRMS08,DRHV10}.

\begin{figure}
\includegraphics[width=6.2in]{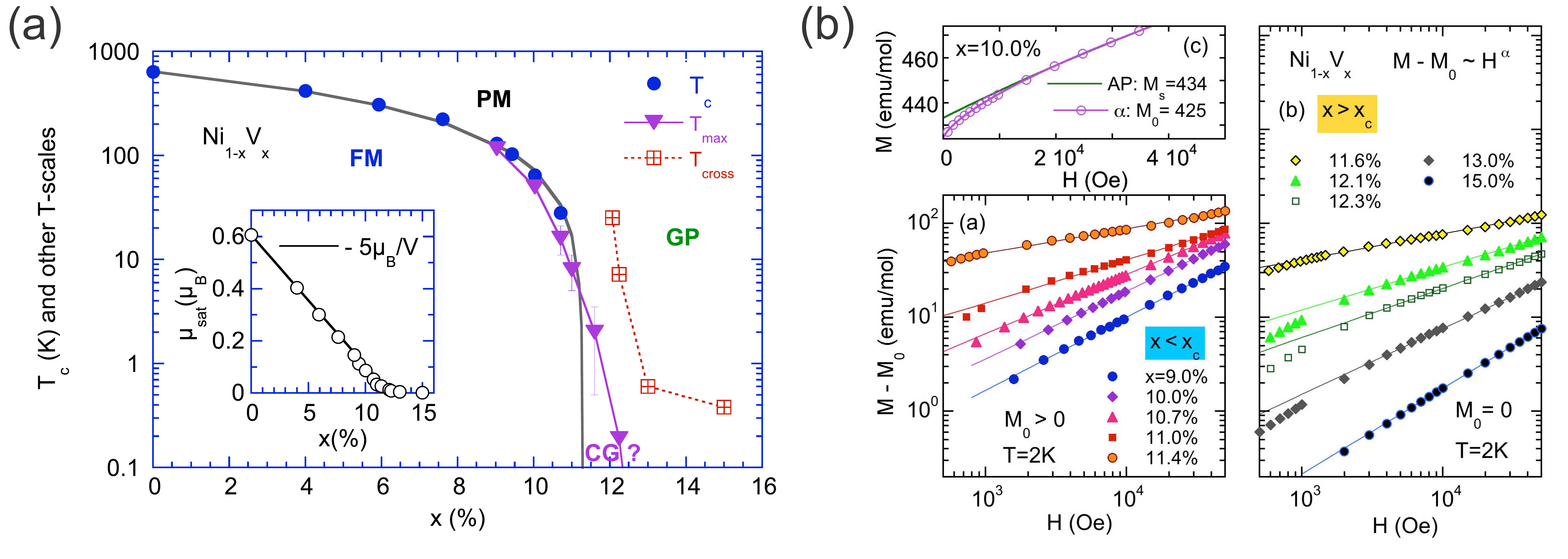}
\caption{(a) Phase diagram of Ni$_{1-x}$V$_x$. FM and PM denote the ferromagnetic and paramagnetic
phases. The (disordered) Griffiths phase (GP) emerges at low temperatures and
$x$ slightly above $x_c$. At the lowest temperatures and close to $x_c$, there may
be a cluster glass (CG) phase (from Ref.\ \cite{UbaidKassisVojtaSchroeder10}.
(b) Low-temperature magnetization-field curves of Ni$_{1-x}$V$_x$ on both sides of the quantum
phase transition. For $x>x_c$, they can be fitted with $M \sim H^\alpha$ with $\alpha=d/z'$ the
Griffiths exponent. For $x<x_c$ they behaves as $M-M_0 \sim H^\alpha$ where $M_0$ is the
spontaneous magnetization (from Ref.\ \cite{Wangetal17}). }
\label{fig:NiV}
\end{figure}

Perhaps the most convincing experimental example of a (class B) quantum Griffiths phase
has been found in the random alloy Ni$_{1-x}$V$_x$. Nickel is a ferromagnet with a
Curie temperature of 627 K. Alloying with vanadium quickly suppresses the ferromagnetism
leading to a quantum phase transition to paramagnetism at a critical vanadium concentration
$x_c$ between 11\% and 12\% (see Fig.\ \ref{fig:NiV}a).
Ubaid-Kassis et al.\ \cite{UbaidKassisVojtaSchroeder10} identified
a Griffiths phase on the paramagnetic side of the quantum phase
transition ($x > x_c$) that shows the predicted power-law behaviors of the susceptibility,
$\chi(T)\sim T^{d/z'-1}$, and the magnetization-field curves, $M(H) \sim H^{d/z'}$
(see Fig.\ \ref{fig:NiV}b). More recently, Wang et al.\ \cite{Wangetal17} discovered a corresponding
Griffiths phase inside the ferromagnetic phase.

Several other examples of magnetic quantum Griffiths phases in metallic systems have been
found in recent years (see Refs.\ \cite{Vojta10,Vojta14} and references therein).
In 2015, Xing et al. \cite{Xingetal15} reported
Griffiths singularities near the superconductor-metal transition in Ga thin films.
Moreover, the elusive ``sliding'' Griffiths phase predicted to occur in layered
superfluids \cite{MGNTV10,PekkerRefaelDemler10} may have been observed in a system
of ultracold atoms.

For quantum phase transitions in class C of the classification,
the quantum Griffiths phase is replaced by a tail of the conventional long-range ordered phase
because the dynamics of sufficiently large rare regions freezes  at zero temperature \cite{Vojta03a}.
The question whether or not Griffiths singularities can be observed at elevated temperatures
has been discussed controversially in the literature
\cite{CastroNetoCastillaJones98,CastroNetoJones00,MillisMorrSchmalian01,MillisMorrSchmalian02}.
Evidence for a smeared quantum phase transition was found in
Sr$_{1-x}$Ca$_x$RuO$_3$ thin films \cite{Demkoetal12}. Pure SrRuO$_3$ is ferromagnetic whereas
CaRuO$_3$ is paramagnetic. The dependence of the critical temperature as well as the magnetization
on the Ca concentration $x$ agree well with the smeared phase transition scenario for itinerant Ising magnets,
adapted to the case of composition-tuning \cite{HrahshehNozadzeVojta11,SNHV12}.

\section{CONCLUSIONS AND OUTLOOK}

In conclusion, we have reviewed the stability of phases and phase transitions
in many-body systems against impurities, defects and other types of quenched disorder.
We have focused on the physics on large length scales where even weak disorder
can lead to qualitative changes of phases and transitions.

A general theme has
emerged from this discussion: When disorder is destroying a long-range
ordered phase or a clean phase transition, exotic new states of
matter are likely to appear that are interesting in there own rights and do not have clean
counterparts. In the following, we summarize the main points and list a few open issues.

\begin{summary}[SUMMARY POINTS]
\begin{enumerate}
\item Random-field disorder prevents spontaneous symmetry breaking in $d\le 2$ for
      discrete order parameter symmetry and in $d\le 4$ for continuous symmetry.
\item Random-field disorder arises naturally for order parameters that break
      real-space symmetries.
\item If long-range order is destroyed by random fields, exotic ``glassy'' phases
      such as the Bragg, spin-density-wave, and pair-density-wave glasses can emerge.
\item Weak random-$T_c$ disorder does not prevent spontaneous symmetry breaking, but it can destroy
      first-order phase transitions and destabilize clean critical points.
\item Critical points in disordered systems feature unconventional scaling scenarios
      that can be classified according to the rare-region dimensionality and the Harris criterion.
\item Exotic Griffiths phases emerge near disordered critical points, including the
      Mott and Bose glasses, the itinerant ferromagnetic quantum Griffiths phase,
      and the sliding phase in layered superfluids.
\end{enumerate}
\end{summary}

\begin{issues}[FUTURE ISSUES]
\begin{enumerate}
\item While the thermodynamics of many of these exotic phenomena is well understood,
      much less is known about the real-time dynamics and transport properties.
\item Theory cannot yet explain the transport properties near disordered quantum phase
      transitions in metallic systems.
\item How does disorder interact with phases characterized by several intertwined orders?
      Does it promote or hinder the formation of vestigial orders?
\item What are the effects of disorder on phases and phase transitions that do not follow
      Landau's paradigm?
\end{enumerate}
\end{issues}

\section*{DISCLOSURE STATEMENT}

The author is not aware of any affiliations, memberships, funding, or financial holdings that
might be perceived as affecting the objectivity of this review.

\section*{ACKNOWLEDGMENTS}

This work would have been impossible without discussions with many friends including
D. Arovas, D. Belitz, A. Castro-Neto, M. Brando, P. Coleman, A. Chubukov,
R. Dickman, V. Dobrosavljevic, R. Fernandes, P. Gegenwart, P. Goldbart, M. Greven,
S. Haas, J.A. Hoyos, F. Igloi, I. K\'eszm\'arki, T.R. Kirkpatrick,
A. del Maestro, A. Millis, E. Miranda, D. Morr, R. Narayanan, G. Refael,
H. Rieger, B. Rosenow, S. Rowley,
S. Sachdev, A. Sandvik, J. Schmalian, A. Schroeder, J. Scott,
J. Toner, N. Trivedi, M. Vojta, X. Wan, and A.P. Young.

This work was supported in part by the NSF under Grant Nos.\ PHY-1125915
and DMR-1506152. T.V. is grateful for the hospitality of the Kavli
Institute for Theoretical Physics, Santa Barbara where part of the work
was performed.


\bibliographystyle{ar-style4}
\bibliography{../00bibtex/rareregions}

\end{document}